\documentclass[letterpaper,10pt,conference]{ieeeconf}

\usepackage{amsmath,amssymb}
\usepackage{graphicx}
\usepackage{booktabs}
\makeatletter
\let\NAT@parse\undefined
\makeatother
\usepackage[authoryear,round]{natbib}
\usepackage{url}
\usepackage{balance}
\usepackage{xcolor}

\graphicspath{{results/}}
\IEEEoverridecommandlockouts
\title{System Identification and acados-Based NMPC for Swing-Up Control of an Underactuated Double Pendulum}

\author{Sichen Li, Venkateswarlu Reddy Konkala, Xiaojie Ning, and Abhishek Alaya Udupa %
\thanks{The first two authors contributed equally to this work. The authors are with Chalmers University of Technology, Gothenburg, Sweden.
Emails: alayaa@chalmers.se, sichenl@chalmers.se, konkala@chalmers.se, and xiaojie@chalmers.se.}%
}

\begin{document}
\maketitle

\begin{abstract}
We identify a base-parameter model of CloudPendulum cell 203 and track an offline swing-up reference with acados SQP-RTI NMPC at a target rate of 400 Hz. A 5 mNm passive-joint assist enabled development-stage swing-up and recovery. In organizer-run testing (16 trials of 300 s per configuration on cells 201--204), assisted pendubot and acrobot mean uptime scores were 80.19 s and 74.61 s. Acrobot scored zero on two cells, and seven trials ended on safety-limit exceptions. Because the assist is prohibited in evaluation, these results are diagnostic rather than qualification scores.
\end{abstract}


\section{Introduction}
CloudPendulum exposes physical acrobot and pendubot cells through a network API for RealAIGym and AI Olympics benchmarks~\citep{kumar2025,wiebe2024,wiebe2022,wiebe2024lessons}. Prior competitions progressed from reinforcement-learning baselines~\citep{wiebe2025second} to real-time acados NMPC~\citep{burchard2025}, with the third competition's overall results summarized by \citet{stark2025third}. The task is single-actuator swing-up and stabilization~\citep{spong1995,spong1995pendubot}, complicated by cell-specific dynamics, friction, constraints, latency, and limited commissioning time.

We contribute a cell-203 identification/validation workflow, a 400-Hz-target SQP-RTI implementation for both configurations, and analysis of development logs and a four-cell organizer benchmark. The novelty is the hardware implementation and failure analysis, not a new NMPC algorithm.

\section{System Description and Constraints}
The state is $x=[q_1,q_2,\dot q_1,\dot q_2]^\top$ (shoulder angle from hanging and relative elbow angle), the input is $u=[\tau_1,\tau_2]^\top$, and $x^\star=[\pi,0,0,0]^\top$. Acrobot drives the elbow; pendubot drives the shoulder. Table~\ref{tab:limits} lists limits and cell-203 settings.

\begin{table}[t]
\caption{Platform limits vs.\ deployed settings.}
\label{tab:limits}
\centering
\scriptsize\setlength{\tabcolsep}{2pt}
\begin{tabular}{lll}
\toprule
Quantity & Platform limit & Deployed\\
\midrule
Torque (measured) & $0.15$\,N\,m & main $\le 0.13$\,N\,m\\
Passive-joint assist & forbidden & $\le 5$\,mNm (development only)\\
Velocity (measured) & $50$\,rad/s & ref.\ $\le 20$; nominal peak $35.3$\\
Control rate & $\le 500$\,Hz & target $400$; mean $383$--$387$\,Hz\\
Development booking & $\le 60$\,s & $10$--$18$\,s control windows\\
\bottomrule
\end{tabular}
\end{table}

\section{System Identification}
\subsection{Data Collection}
Cell-203 data were collected at 200 Hz with $|\tau|\le60$ mNm and measured time stamps (Table~\ref{tab:data}). Free swings excite gravity, inertia, and passive friction; two-phase D-optimal multisines excite coupling and actuated friction; guarded chirps cover higher speeds. The pilots used a conservative local abort near 40 rad/s, not the current 50 rad/s platform limit. Of 19 attempts, 16 were usable. Holding out one 6-s free swing and one 3-s chirp left 10,586 fitting transitions.

\begin{table}[t]
\caption{Identification data (cell 203, 200 Hz); aborts are local.}
\label{tab:data}
\centering
\footnotesize\setlength{\tabcolsep}{4pt}
\begin{tabular}{lcccc}
\toprule
Trajectory set & Runs & Length & Peak $|\dot q|$ & Aborts\\
 & & (s) & (rad/s) & \\
\midrule
Free-swing, $q_1(0){\in}[0.6,2.4]$\,rad & 5 & $\le 6.0$ & 4.7--37.5 & 1\\
D-optimal multisine (A/B) & 2 & 4.0 & 6.2--11.9 & 0\\
High-speed pilots (unguarded) & 2 & $\le 0.8$ & 39.4--39.9 & 2\\
Burst chirps (velocity guard) & 10 & 3.0 & 11.5--20.2 & 0\\
\bottomrule
\end{tabular}
\end{table}

\subsection{Identified Model}
Because individual link parameters $(m_i,l_i,r_i,I_i)$ are not separately identifiable, we use the standard base-parameter model~\citep{wiebe2024}:
\begin{equation}
M(q)\ddot q+h(q,\dot q)+G(q)+\tau_f(\dot q)=\tau,
\end{equation}
where $h=[-P_2\sin q_2(2\dot q_1\dot q_2+\dot q_2^2),\;P_2\sin q_2\dot q_1^2]^\top$ and
\begin{equation}
\begin{aligned}
M_{11} &= P_1 + 2P_2\cos q_2, \quad M_{12} = P_3 + P_2\cos q_2,\\
M_{21} &= M_{12}, \quad M_{22} = P_6,\\
G_1 &= P_4\sin q_1 + P_5\sin(q_1{+}q_2),\\
G_2 &= P_5\sin(q_1{+}q_2),
\end{aligned}
\end{equation}
where independent $P_6$ absorbs elbow-axis motor/rotor inertia. Per-joint friction is
\begin{equation}
\begin{aligned}
\tau_{f,i}={}&b_i\dot q_i+c_{f,i}\tanh(\dot q_i/\varepsilon)
 +d_i\dot q_i\sqrt{\dot q_i^2+v_0^2},\\
&\varepsilon=0.05\,\mathrm{rad/s},\qquad v_0=0.5\,\mathrm{rad/s}.
\end{aligned}
\end{equation}
The first two terms are viscous and smoothed Coulomb friction~\citep{armstrong1994}; the last is a differentiable quadratic-drag test retained to reproduce the deployed model, although its fit is negligible. Twelve parameters were fitted by 80-step, re-anchored RK4 simulation-error minimization using measured $\Delta t$, IPOPT, and a trust-region least-squares fallback~\citep{wachter2006,andersson2019}. Table~\ref{tab:params} gives the result. A normalized Fisher-information condition number of $2.2{\times}10^2$ flags $c_{f1}$ and $d_1$ as weakly identifiable; it is not a confidence interval.

\begin{table}[t]
\caption{Cell-203 base parameters. Units: $P_{1,2,3,6}$ in kg\,m$^2$; $P_{4,5},c_f$ in N\,m; $b$ in N\,m\,s/rad; $d$ in N\,m\,s$^2$/rad$^2$.}
\label{tab:params}
\centering
\begin{tabular}{lc@{\hskip 12pt}lc@{\hskip 12pt}lc}
\toprule
$P_1$ & $6.50\,\mathrm{e}{-4}$ & $P_4$ & $9.07\,\mathrm{e}{-2}$ & $c_{f1}$ & $1.72\,\mathrm{e}{-3}$\\
$P_2$ & $1.75\,\mathrm{e}{-4}$ & $P_5$ & $3.16\,\mathrm{e}{-2}$ & $c_{f2}$ & $3.91\,\mathrm{e}{-3}$\\
$P_3$ & $1.84\,\mathrm{e}{-4}$ & $b_1$ & $2.81\,\mathrm{e}{-4}$ & $d_1$ & $\approx 0$\\
$P_6$ & $2.06\,\mathrm{e}{-4}$ & $b_2$ & $1.04\,\mathrm{e}{-6}$ & $d_2$ & $7.9\,\mathrm{e}{-7}$\\
\bottomrule
\end{tabular}
\end{table}

\subsection{Validation Metrics}
\label{subsec:validation_metrics}
\subsubsection{Single-Step Prediction Baseline}
RK4 propagation of measured $(x[k],\tau[k],\Delta t)$ gives $\hat x[k+1]$. Absolute component errors are pooled over samples and joints, with wrapped angle differences. Fitting-set angle median/p95 and velocity p95 are $0.070^\circ/0.638^\circ$ and $0.639$ rad/s; on the two validation trajectories (1,798 transitions), they are $0.114^\circ/0.795^\circ$ and $0.670$ rad/s. These trajectories select the horizon and are not an untouched test set. Identification and in-run errors use approximately 5-ms and 2.6-ms steps, respectively, and are not directly comparable.

\subsubsection{Rationale for NMPC Prediction Horizon}
\begin{figure}[t]
\centering
\includegraphics[width=\columnwidth]{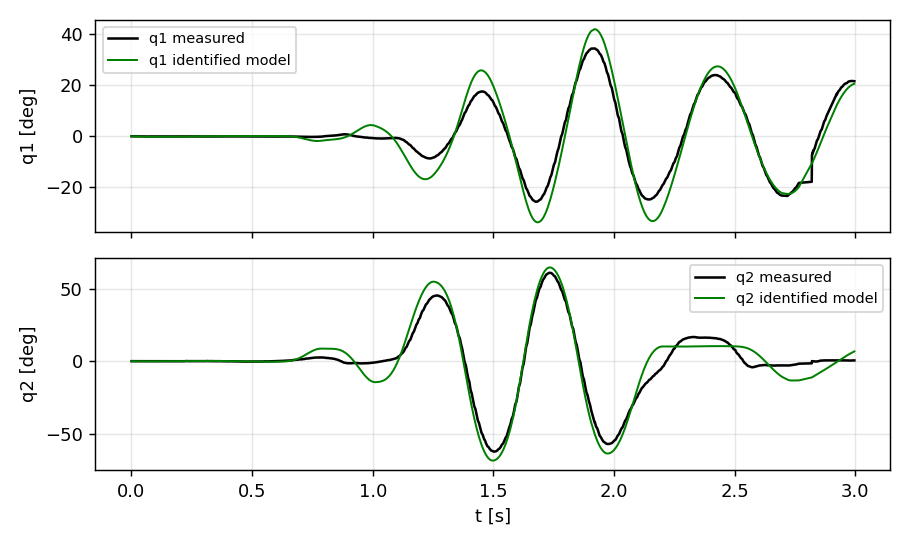}
\caption{Validation on the held-out high-speed excitation \texttt{oid\_hi\_long\_09}: measured angles (black) vs.\ open-loop simulation with the identified model (green).}
\label{fig:sysid}
\end{figure}

Angle-error p95 at 0.20, 0.35, 0.40, and 0.60 s is $8.49^\circ$, $11.68^\circ$, $12.07^\circ$, and $14.01^\circ$. We choose $N_{\text{MPC}}=18$ (0.36 s) between the tested 0.35--0.40 s points; $12^\circ$ is not a hard boundary because RTI re-anchors each update. Full-rollout $q_1/q_2$ RMSE is $8.0^\circ/10.9^\circ$ for the 6-s free swing and $4.3^\circ/6.6^\circ$ for the chirp (Fig.~\ref{fig:sysid}).

\subsubsection{State Estimation and Command Execution}
\begin{itemize}
    \item \textbf{Commands:} NPZ logs contain solver output and safety-clipped sent commands, not measured torque; Figs.~\ref{fig:pend-run}--\ref{fig:acro-run} therefore show command execution, not torque tracking.
    \item \textbf{State:} Velocity fuses wrapped position differences and hardware velocity using a complementary filter ($w_{\text{pos}}=0.70$), 3-sample median, and EMA ($\alpha=0.55$). After 5 s, worst-joint velocity/angle tracking p95 is $1.60$ rad/s/$1.12^\circ$ (acrobot) and $1.14$ rad/s/$0.77^\circ$ (pendubot).
    \item \textbf{Timing:} Loop periods are $2.583\pm0.988$ ms and $2.612\pm0.959$ ms for acrobot and pendubot.
\end{itemize}

\section{Control Method}
\label{sec:control_method}

\subsection{Offline Trajectory Generation}
IPOPT solves the following multiple-shooting optimal control problem (OCP), with $U_k=[\tau_{1,k},\tau_{2,k}]^\top$, $T=3.0$ s, $N=150$, and $\Delta t_{\text{ref}}=20$ ms:
\begin{align}
\min_{\mathbf{X}, \mathbf{U}} \quad & \sum_{k=0}^{N-1} U_k^\top R_{\mathrm{ref}}U_k \, \Delta t + 50\| X_N - x^* \|_2^2 \label{eq:ocp_cost} \\
\text{s.t.} \quad & X_0 = [0, 0, 0, 0]^\top, \nonumber \\
& X_{k+1} = f_{\text{RK4}}(X_k, U_k, \Delta t), \nonumber \\
& -\bar U \le U_k \le \bar U, \quad |\dot{q}_{i,k}| \le 20\,\mathrm{rad/s}, \nonumber\\
& \cos q_{1,N}=-1,\quad\sin q_{1,N}=0,\quad q_{2,N}=0, \nonumber\\
& \dot q_{1,N}=\dot q_{2,N}=0. \nonumber
\end{align}
For acrobot, $\bar U=[0.005,0.13]^\top$ N m and $R_{\mathrm{ref}}=\mathrm{diag}(5,1)$; for pendubot, $\bar U=[0.13,0.005]^\top$ N m and $R_{\mathrm{ref}}=I$. Terminal constraints impose upright, the terminal cost selects the $q_1\approx\pi$ branch, and a 3-s upright hold is appended.

\subsection{Online NMPC Formulation}
An \texttt{acados} SQP-RTI solver~\citep{verschueren2022} tracks $(X_{\text{ref}},U_{\text{ref}})$ over $N_{\text{MPC}}=18$ ($T_{\text{pred}}=0.36$ s):

\begin{subequations}
\label{eq:nmpc_formulation}
\begin{align}
\min_{\mathbf{x}, \mathbf{u}, \mathbf{v}} \quad & \frac{1}{2}\sum_{k=0}^{N_{\text{MPC}}-1} \left( \| x_k - x_{\text{ref},k} \|_Q^2 + \|u_k-u_{\text{ref},k}\|_R^2 \right) \nonumber \\
& \quad + \frac{1}{2}\| x_{N_{\text{MPC}}} - x_{\text{ref},N_{\text{MPC}}} \|_{Q_f}^2 + L_{\text{slack}}(v) \label{eq:nmpc_cost} \\
\text{s.t.} \quad & x_0 = x_{\text{meas}}, \label{eq:nmpc_init} \\
& x_{k+1} = f_{\text{RK4}}(x_k, u_k, \Delta t_{\text{ref}}), \quad k = 0, \dots, N_{\text{MPC}}-1, \label{eq:nmpc_dyn} \\
& -\bar U \le u_k \le \bar U, \label{eq:nmpc_u_bound} \\
& -\dot{q}_{\max} - v_{l,k} \le \dot{q}_{i,k} \le \dot{q}_{\max} + v_{u,k}, \quad v \ge 0, \label{eq:nmpc_v_bound}
\end{align}
\end{subequations}

Here $\dot q_{\max}=20\,\mathrm{rad/s}$, $v=\{v_{l,i,k},v_{u,i,k}\}$, and the implemented acados slack convention is
\begin{equation}
L_{\mathrm{slack}}=\sum_{k,i}\left[\frac{Z}{2}(v_{l,i,k}^2+v_{u,i,k}^2)+z(v_{l,i,k}+v_{u,i,k})\right],
\end{equation}
with $Z=10^5$, $z=10^3$, $Q=\mathrm{diag}(120,50,2,2)$, and $Q_f=\mathrm{diag}(3000,1200,80,80)$. Acrobot and pendubot use $R=\mathrm{diag}(0.1,0.05)$ and $\mathrm{diag}(0.05,0.1)$. The deployed solver uses ordinary differences of wrapped hardware angles, creating a possible $-\pi/\pi$ branch-cut error; only reported errors are wrapped.

The 20-ms prediction grid is distinct from the 2.5-ms target update: reference index $\lfloor t_{\text{meas}}/\Delta t_{\text{ref}}\rfloor$ advances through stored nodes, and only $u_0$ is applied. Each call uses one four-stage ERK step per node, one SQP-RTI iteration, Partial Condensing HPIPM, and a Gauss--Newton Hessian. Mean rates are 387 Hz (acrobot) and 383 Hz (pendubot); nonzero solver status yields a clipped zero command.

Pendubot also constrains $u_{0,j}$ to $u_{\mathrm{prev},j}\pm\rho_j\Delta t_{\mathrm{apply}}$, intersected with torque bounds, where $\rho=[20,100]^\top$ N m/s (shoulder/elbow). LQR and speed braking were disabled.

\subsection{Micro-Compensation Strategy}
A 5 mNm passive-joint channel (3.8\% of the main limit), comparable to $c_{f1}=1.72$ mNm and $c_{f2}=3.91$ mNm, was enabled during development. It is prohibited in evaluation, so assisted runs are diagnostic only.

\subsection{Disturbance Tests}
Development tests applied a position jump to $[\pi-0.25,0.25]$ rad at 5 s and, for pendubot, $\pm15$ mNm passive-joint pulses at 7 and 9 s. They are smaller than the organizer protocol in Section~\ref{sec:official_benchmarking}.

\section{Experimental Results}
\subsection{Nominal Swing-Up and Stabilization}
\begin{figure*}[t]
\centering
\includegraphics[width=0.80\textwidth]{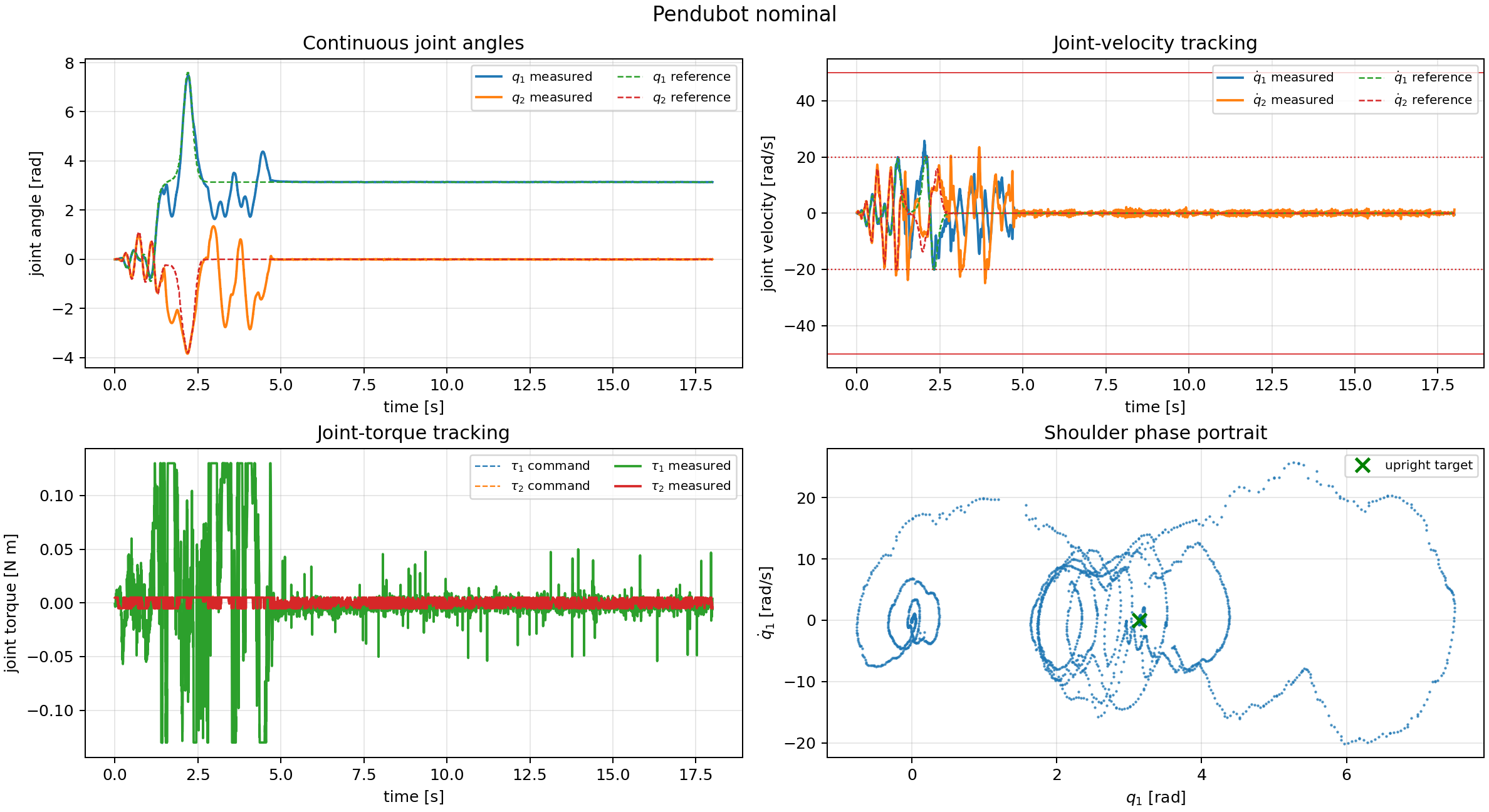}
\caption{Assisted pendubot run (cell 203, session 2275056600049151232): continuous angles, filtered velocity, NMPC and sent commands, and shoulder phase portrait. Torque was not measured.}
\label{fig:pend-run}
\end{figure*}

\begin{figure*}[t]
\centering
\includegraphics[width=0.80\textwidth]{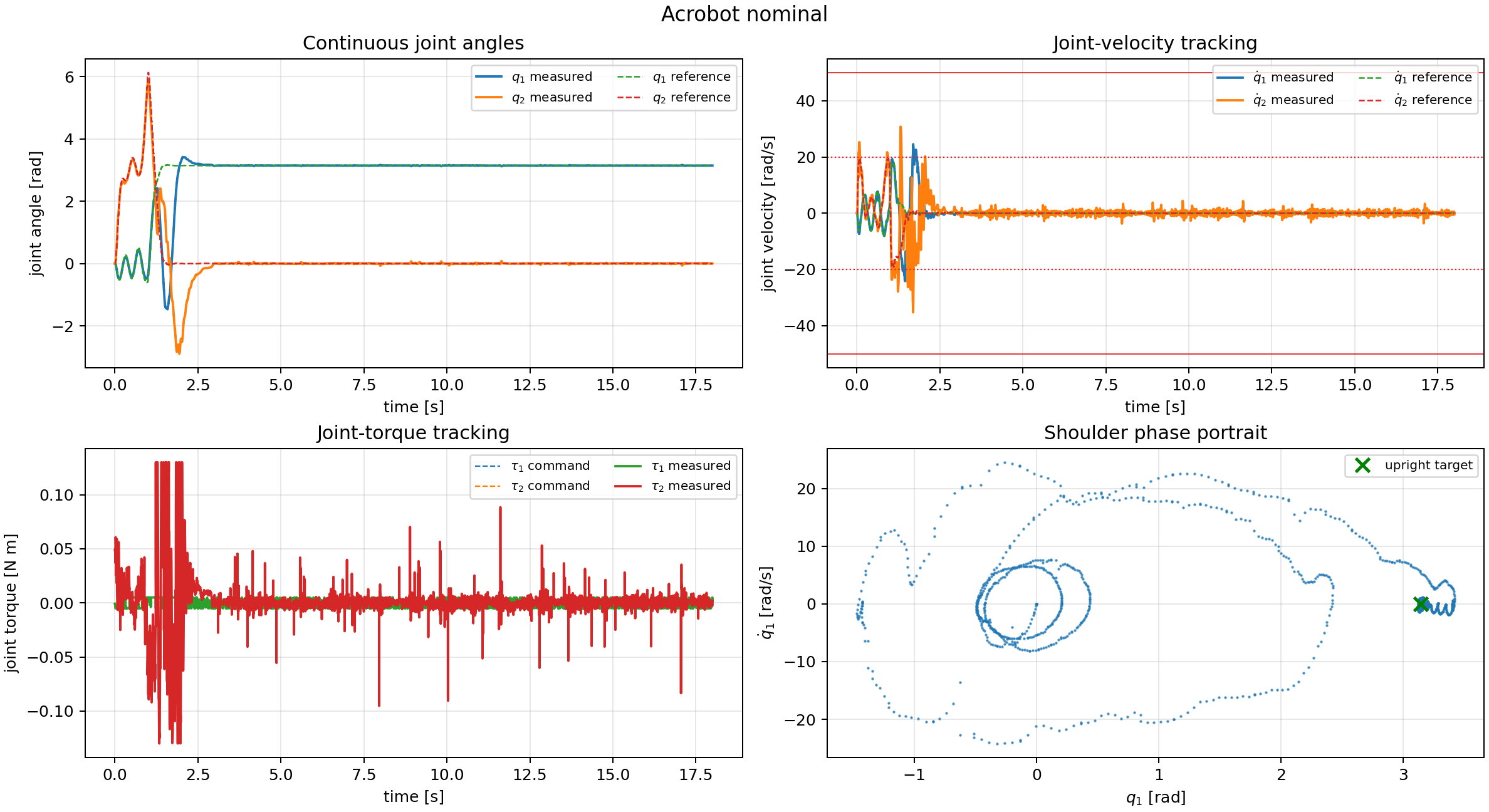}
\caption{Assisted acrobot run (cell 203, session 449575849258851072); conventions as in Fig.~\ref{fig:pend-run}.}
\label{fig:acro-run}
\end{figure*}

Table~\ref{tab:nominal} and Figs.~\ref{fig:pend-run}--\ref{fig:acro-run} report one retained $\sim18$-s assisted log per configuration. Both reached upright without aborting; first arrival was approximately 2.5 s (acrobot) and 4.6 s (pendubot). These selected runs do not estimate success probability.

\begin{table}[t]
\caption{Selected assisted cell-203 runs (one per configuration).}
\label{tab:nominal}
\centering
\scriptsize\setlength{\tabcolsep}{2pt}
\begin{tabular}{lcc}
\toprule
 & Acrobot & Pendubot\\
\midrule
Loop steps / aborted & 6970 / no & 6892 / no\\
Final $|q_1-\pi|$ (rad) & 0.00134 & 0.00249\\
Peak $|\dot q|$ (rad/s) & 35.3 & 25.8\\
Main sent command (N\,m) & 0.130 (elbow) & 0.130 (shoulder)\\
Passive sent command (N\,m) & 0.005 (shoulder) & 0.005 (elbow)\\
Failed solves & 0/6970 & 9/6892\\
1-step $q$ err p95 (rad) & 0.00580 & 0.00738\\
1-step $\dot q$ err p95 (rad/s) & 0.642 & 0.596\\
Solve mean$\pm$SD / p99 (ms) & 0.481$\pm$0.325 / 1.272 & 0.540$\pm$0.497 / 1.506\\
Loop period mean$\pm$SD (ms) & 2.583$\pm$0.988 & 2.612$\pm$0.959\\
\bottomrule
\end{tabular}
\end{table}

\subsection{Disturbance Recovery}
All three assisted runs ended upright without aborting (Table~\ref{tab:disturbance}). Errors cover each full run; recovery time is unavailable. The acrobot jump peaked at 46 rad/s filtered (42 raw), below 50 rad/s. Its torque-disturbance trial lacks a numerical log and is excluded.

\begin{table}[t]
\caption{Disturbance-recovery runs (all non-aborting).}
\label{tab:disturbance}
\centering
\scriptsize\setlength{\tabcolsep}{2pt}
\begin{tabular}{lccc}
\toprule
 & \multicolumn{1}{c}{Acrobot} & \multicolumn{2}{c}{Pendubot}\\
\cmidrule(lr){2-2}\cmidrule(lr){3-4}
 & pos.\ jump & torque pulses & pos.\ jump\\
\midrule
Final $|q_1-\pi|$ (rad) & 0.0026 & 0.0017 & 0.0035\\
Peak $|\dot q|$ (rad/s) & 46.0 & 26.3 & 31.9\\
Peak main sent command (N\,m) & 0.130 & 0.130 & 0.130\\
1-step $q$ p95 (rad) & 0.0159 & 0.0143 & 0.0065\\
1-step $\dot q$ p95 (rad/s) & 2.42 & 0.90 & 0.54\\
Failed solves & 3/3590 & 0/6484 & 10/6637\\
Mean loop period (ms) & 2.79 & 2.78 & 2.71\\
\bottomrule
\end{tabular}
\end{table}

\section{Official Benchmarking}
\label{sec:official_benchmarking}
Organizers ran four 300-s disturbance trials per cell and configuration on cells 201--204, separate from the development booking limit. Table~\ref{tab:benchmark} treats their \texttt{score} as uptime and includes safety-terminated trials.

\begin{table}[t]
\caption{Organizer assisted benchmark: four 300-s trials per cell/configuration; scores in seconds.}
\label{tab:benchmark}
\centering
\scriptsize\setlength{\tabcolsep}{3pt}
\begin{tabular}{lcc}
\toprule
 & Acrobot & Pendubot\\
\midrule
All-cell mean$\pm$sample SD & $74.61\pm78.64$ & $80.19\pm28.73$\\
All-cell range & 0--178.06 & 26.12--134.58\\
Cell 201 mean & 0.00 & 65.55\\
Cell 202 mean & 150.22 & 87.90\\
Cell 203 mean & 148.20 & 103.67\\
Cell 204 mean & 0.00 & 63.64\\
Zero-score trials & 8 / 16 & 0 / 16\\
Organizer status SUCCESS / FAILED & 12 / 4 & 13 / 3\\
Velocity-limit exceptions & 4 & 2\\
Torque-limit exceptions & 0 & 1\\
Mean fraction of 300\,s & 24.9\% & 26.7\%\\
\bottomrule
\end{tabular}
\end{table}

\texttt{SUCCESS}/\texttt{FAILED} denotes exception status, not task success. Six trials exceeded 50 rad/s (51.93--55.09 rad/s); one pendubot trial reported $-0.301$ N m measured torque, which cannot be separated into command, disturbance, and response. The 5 mNm assist was active. Acrobot scored zero in all cell-201/204 trials but about 150 s on cells 202/203; aggregate performance is therefore cell dependent. Organizer files omit the loaded model/reference provenance, so cross-cell model transfer is unverified.

\section{Discussion}\label{sec:discussion}
Four uncompensated pendubot runs had $0.065^\circ$ median one-step error before leaving the reference at 0.30--0.35 s; three exceeded $30^\circ$ path error at 0.866--0.875 s. No prior saturation was observed. Gross local-model or torque-scale error is therefore less likely, but finite-horizon error, estimation, and trackability remain possible. The 5 mNm assist is comparable to identified Coulomb friction, but assisted and uncompensated runs are not a controlled ablation.

The uncompensated 80 mNm reference contains a 200-ms torque flat-top at 0.34--0.52 s, overlapping departure. Re-optimization at 130 mNm removes it with nearly unchanged $\int\tau^2dt$ (0.00439 to 0.00434); causality remains untested.

\section{Limitations and Future Work}
Open issues are evaluation-compliant control, a comparable baseline, unseen-cell commissioning, event-aligned/provenance-complete logs, a periodic angle residual, and constraint-active references.

\section{Conclusion}
The identified cell-203 model supported assisted NMPC swing-up and recovery in both configurations. Organizer scores were 80.19 s (pendubot) and 74.61 s (acrobot), with strong cell dependence. Because passive assist is prohibited, qualification performance remains unproven; uncompensated tests require further reference and robustness development.

\balance


\begin{thebibliography}{99}

\bibitem[Andersson et~al.(2019)]{andersson2019}
Andersson, J.A.E., Gillis, J., Horn, G., Rawlings, J.B. and Diehl, M. (2019) `CasADi: A software framework for nonlinear optimization and optimal control', \emph{Mathematical Programming Computation}, 11(1), pp. 1--36. doi: 10.1007/s12532-018-0139-4.

\bibitem[Armstrong-H\'elouvry et~al.(1994)]{armstrong1994}
Armstrong-H\'elouvry, B., Dupont, P. and Canudas de Wit, C. (1994) `A survey of models, analysis tools and compensation methods for the control of machines with friction', \emph{Automatica}, 30(7), pp. 1083--1138. doi: 10.1016/0005-1098(94)90209-7.

\bibitem[Burchard and Stark(2025)]{burchard2025}
Burchard, B. and Stark, F. (2025) `Real-time model predictive control for the swing-up problem of an underactuated double pendulum', \emph{arXiv preprint}, arXiv:2504.05363. doi: 10.48550/arXiv.2504.05363.

\bibitem[Kumar(2025)]{kumar2025}
Kumar, S. (2025) `Swinging pendulums on the cloud: Digitalization of simulation \& experimental infrastructure for feedback-based active learning', in \emph{Chalmers Konferens om Undervisning och L\"arande (KUL 2025)}, Gothenburg, Sweden.

\bibitem[Spong(1995)]{spong1995}
Spong, M.W. (1995) `The swing up control problem for the Acrobot', \emph{IEEE Control Systems Magazine}, 15(1), pp. 49--55. doi: 10.1109/37.341864.

\bibitem[Spong and Block(1995)]{spong1995pendubot}
Spong, M.W. and Block, D.J. (1995) `The Pendubot: A mechatronic system for control research and education', in \emph{Proceedings of the 34th IEEE Conference on Decision and Control}, New Orleans, Louisiana, USA, vol. 1, pp. 555--556. doi: 10.1109/CDC.1995.478951.

\bibitem[Stark et~al.(2025)]{stark2025third}
Stark, F., Wiebe, F., Burchard, B., Choe, J.S.B., Choi, B., Kim, J.-K., Turcato, N., Cal\`i, M., Dalla Libera, A., Giacomuzzo, G., Carli, R., Romeres, D., Alentev, I., Domrachev, I., Kozlov, L., Frey, J., Nurkanovi\'c, A., Vyas, S., Mronga, D., Kirchner, F. and Kumar, S. (2025) `Towards global swing-up policies for the underactuated double pendulum: Results from the 3rd AI Olympics with RealAIGym competition', \emph{IEEE Control Systems Magazine}, manuscript under review.

\bibitem[Verschueren et~al.(2022)]{verschueren2022}
Verschueren, R., Frison, G., Kouzoupis, D., Frey, J., van Duijkeren, N., Zanelli, A., Novoselnik, B., Albin, T., Quirynen, R. and Diehl, M. (2022) `acados---a modular open-source framework for fast embedded optimal control', \emph{Mathematical Programming Computation}, 14(1), pp. 147--183. doi: 10.1007/s12532-021-00208-8.

\bibitem[W\"achter and Biegler(2006)]{wachter2006}
W\"achter, A. and Biegler, L.T. (2006) `On the implementation of an interior-point filter line-search algorithm for large-scale nonlinear programming', \emph{Mathematical Programming}, 106(1), pp. 25--57. doi: 10.1007/s10107-004-0559-y.

\bibitem[Wiebe et~al.(2022)]{wiebe2022}
Wiebe, F., Vyas, S., Maywald, L.J., Kumar, S. and Kirchner, F. (2022) `RealAIGym: Education and research platform for studying athletic intelligence', in \emph{Proceedings of the Robotics: Science and Systems Workshop Mind the Gap: Opportunities and Challenges in the Transition Between Research and Industry}, New York, New York, USA, 1 July.

\bibitem[Wiebe et~al.(2024a)]{wiebe2024}
Wiebe, F., Kumar, S., Shala, L.J., Vyas, S., Javadi, M. and Kirchner, F. (2024a) `Open source dual-purpose Acrobot and Pendubot platform: Benchmarking control algorithms for underactuated robotics', \emph{IEEE Robotics \& Automation Magazine}, 31(2), pp. 113--124. doi: 10.1109/MRA.2023.3341257.

\bibitem[Wiebe et~al.(2024b)]{wiebe2024lessons}
Wiebe, F., Turcato, N., Dalla Libera, A., Zhang, C., Vincent, T., Vyas, S., Giacomuzzo, G., Carli, R., Romeres, D., Sathuluri, A., Zimmermann, M., Belousov, B., Peters, J., Kirchner, F. and Kumar, S. (2024b) `Reinforcement learning for athletic intelligence: Lessons from the 1st ``AI Olympics with RealAIGym'' competition', in Larson, K. (ed.), \emph{Proceedings of the Thirty-Third International Joint Conference on Artificial Intelligence (IJCAI-24)}. International Joint Conferences on Artificial Intelligence Organization, article 1043, pp. 8833--8837. doi: 10.24963/ijcai.2024/1043.

\bibitem[Wiebe et~al.(2025)]{wiebe2025second}
Wiebe, F., Turcato, N., Dalla Libera, A., Choe, J.S.B., Choi, B., Faust, T.L., Maraqten, H., Aghadavoodi, E., Cali, M., Sinigaglia, A., Giacomuzzo, G., Carli, R., Romeres, D., Kim, J.-K., Susto, G.A., Vyas, S., Mronga, D., Belousov, B., Peters, J., Kirchner, F. and Kumar, S. (2025) `Reinforcement learning for robust athletic intelligence: Lessons learned from the second AI Olympics with RealAIGym competition', \emph{IEEE Robotics \& Automation Magazine}, pp. 2--12. doi: 10.1109/MRA.2025.3631571.

\end{thebibliography}
\end{document}